\documentstyle[12pt,blois]{article}

\include{psfig} \psnoisy 

\begin{document}
\heading{THE FIR/SUBMM WINDOW\\ ON GALAXY FORMATION}

\author{B. Guiderdoni$^1$, F.R. Bouchet$^1$, J. Devriendt$^1$, E. Hivon$^2$, 
\& J.L. Puget$^3$} 
{$^1$Institut d'Astrophysique de Paris, CNRS, 98bis boulevard Arago, F--75014 
Paris}
{$^2$ Canadian Institute for Theoretical Astrophysics, 
University of Toronto, McLennan Labs, 60 St George St., Toronto, 
Ontario M5S 1A1}
{$^3$Institut d'Astrophysique Spatiale, B\^at 121, Universit\'e Paris XI, 
F--91405, Orsay Cedex}
 
\begin{bloisabstract}
Our view on the deep universe has been so far biased towards optically 
bright galaxies. Now,
the measurement of the Cosmic Infrared Background in FIRAS and
DIRBE residuals, and the observations of FIR/submm sources by the ISOPHOT
and SCUBA instruments begin unveiling the ``optically dark side''
of galaxy formation. Though the origin of dust heating is still unsolved, 
it appears very likely that a large fraction of the FIR/submm emission is due 
to heavily---extinguished star formation. Consequently,
the level of the CIRB implies that about 2/3 of galaxy/star formation 
in the universe is hidden by dust shrouds. In this review, we introduce a new
modeling of galaxy formation and evolution that provides us with specific 
predictions in FIR/submm wavebands. These predictions are compared with 
the current status of the observations. Finally, the capabilities of current 
and forthcoming instruments for all--sky and deep surveys of FIR/submm 
sources are briefly described.
\end{bloisabstract}

\section{Introduction}
Recent observational breakthroughs have made possible the measurement
of the Star Formation Rate (SFR) history of the universe from rest--frame 
UV fluxes of moderate-- and high--redshift galaxies (Lilly {\it et al.} 1996,
Madau {\it et al.} 1996, 1998).
The strong peak observed at $z \sim 1.5$ seems to be correlated with
the decrease of the cold--gas comoving density in damped Lyman--$\alpha$ 
systems between $z=2$ and $z=0$ (Lanzetta {\it et al.} 1995,
Storrie--Lombardi {\it et al.} 1996) These results nicely fit in 
a view where star formation in bursts triggered by interaction/merging
consumes and enriches the gas content of galaxies as time goes on.
Such a view is qualitatively predicted within the paradigm of 
hierarchical growth of structures in which galaxy formation is a continuous
process (see e.g. Baugh {\it et al.} 1998).

However, these observational data come from optical
surveys that probe the rest--frame UV and visible emission of 
high--$z$ galaxies. In the early universe, what fraction of star/galaxy 
formation was hidden by dust that absorbs UV/visible starlight and thermally 
re--radiates at larger wavelengths ?
In the {\it local} universe (and thanks to
IRAS and ISO observations), we know that
about 30 \% of the bolometric luminosity of galaxies is 
radiated in the IR (Soifer \& Neugebauer 1991), and that a large fraction of 
dust heating is due to young stellar populations (Genzel {\it et al.} 1998). 
Now, IR/submm observations are beginning to unveil what actually 
happened at higher redshift.

We might have kept so far the prejudice that high--redshift
galaxies have little extinction, simply because their heavy--element 
abundances are low (typically 1/100 to 1/10 of solar at $z>2$). However, low 
abundances do not necessarily mean low extinction. For instance, if we 
assume that dust grains have a size distribution similar to the one of our 
Galaxy 
($n(a)da \propto a^{-3.5}$ with $a_{min} \le a \le a_{max}$), and are
homogeneously distributed in a region with radius $R$, the optical depth varies
as $\tau \propto a_{min}^{-0.5}R$ while the total dust mass varies as  
$M_{dust} \propto a_{max}^{0.5}R^3$. For given dust mass and size 
distribution, there is more extinction where grains are small, and 
close to the heating sources. This is probably the reason why
Thuan {\it et al.} (1998) observed a 
significant dust emission in the extremely metal--poor galaxy SBS0335-052.

In this context, we hereafter briefly review the attempts to correct the 
UV fluxes emitted by high--redshift galaxies for the effect of extinction, 
as well as recent measurements of the ``Cosmic Infrared 
Background'' (hereafter CIRB), and deep surveys at FIR/submm wavelengths.
These observations strongly suggest that a significant fraction of the 
young stellar populations is hidden by dust. Finally, we propose a 
semi-analytic modeling of galaxy formation and evolution in which the 
computation of dust extinction and emission is explicitly implemented.
This model is helpful to prepare forthcoming observations in the FIR/submm 
range.

\begin{figure}[htb]
\centerline{
\psfig{figure=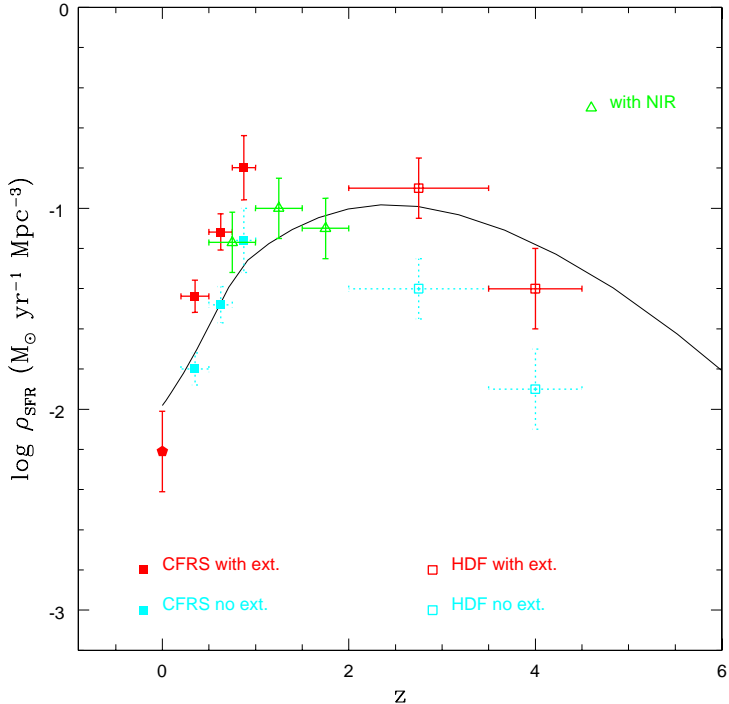,width=0.6\textwidth}
}
\caption{\small The evolution of the 
cosmic Star Formation Rate comoving density $\rho_{SFR}$ with redshift $z$. 
For the Canada--France Redshift Survey ($z<1$), the solid dots and error bars
drawn with dotted lines and solid lines respectively give values
uncorrected for extinction (Lilly {\it et al.} 1996), and 
values estimated from a multi--wavelength analysis including IR,
submm, and radio data (Flores {\it et al.} 1999). For the Hubble Deep Field,
the open squares and error bars drawn with dotted lines and solid lines 
respectively give values uncorrected for extinction (Madau {\it et al.} 
1996, 1998), and values corrected for an average $<E(B-V)>=0.09$ and 
the SMC extinction curve (Pettini {\it et al.} 1998).
The corrections derived by Meurer {\it et al.} (1997) would shift the 
corrected points upwards by $\sim$0.5 dex. Finally, the open triangles 
show values
determined from the HDF by a method with photometric redshifts involving
visible and near--IR photometry (Connolly {\it et al.} 1997). The rest--frame 
UV fluxes are converted into SFRs with the spectrophotometric model 
described in Guiderdoni {\it et al.} (1998), and a Salpeter IMF. 
The solid line shows the best
model in the latter paper (the so--called ``model E'').}
\end{figure}

\section{The issue of extinction in high--redshift galaxies}
Deep spectroscopic surveys and the development of the powerful UV drop--out 
technique have led to the reconstruction of the cosmic SFR 
comoving density
(Lilly {\it et al.} 1996, Steidel \& Hamilton 1993, Steidel {\it et al.} 
1996, 1999, Madau {\it et al.} 1996, 1998). However, a complete assessment
of the effect of extinction on UV fluxes emitted by young
stellar populations, and of the luminosity budget of star--forming galaxies
is still to come.

The cosmic SFR density determined only from UV observations of
the Canada--France Redshift Survey has been recently revisited with a 
multi--wavelength approach including IR, submm, and radio observations. 
The result is an upward correction of the previous values by an 
average factor 2.9 (Flores {\it et al.} 1999). 
At higher redshift, various authors have attempted to estimate
the extinction correction and to recover the fraction of UV starlight 
absorbed by dust (e.g. Meurer {\it et al.} 1997, Pettini {\it et al.} 1998). 
It turns out 
that the observed slope $\alpha$ of the UV spectral energy distribution
$F_\lambda (\lambda) \propto \lambda^\alpha$ (say, around 2200 \AA)
is flatter than the standard value  $\alpha_0 \simeq -2.5$ 
computed from models of spectrophotometric evolution. The derived 
extinction 
corrections are large and differ according to the method. For instance,
Pettini {\it et al.} (1998) and coworkers fit a typical extinction curve 
(the Small Magellanic Cloud one) to the observed colors,
whereas Meurer {\it et al.} (1997) and coworkers use an 
empirical relation between $\alpha$ and the FIR to 2200 \AA\ luminosity ratio
in {\it local} starbursts. The former authors
derive $<E(B-V)> \simeq 0.09$ resulting 
in a factor 2.7 absorption at 1600 \AA, whereas the latter 
derive $<E(B-V)> \simeq 0.30$ resulting in a 
factor 10 absorption. This discrepancy suggests sort of
a bimodal distribution of the young stellar populations : the first 
method would take into account the stars detected in the UV with relatively
moderate reddening/extinction, while the 
second one would phenomenologically add the contributions of these ``apparent''
stars and of heavily--extinguished stars.
 
Fig. 1 shows the cosmic SFR comoving density in the early version
(no extinction), and after the work by Flores {\it et al.} (1999) 
at $z<1$ and extinction corrections. 

\begin{figure}[htb]
\centerline{
\psfig{figure=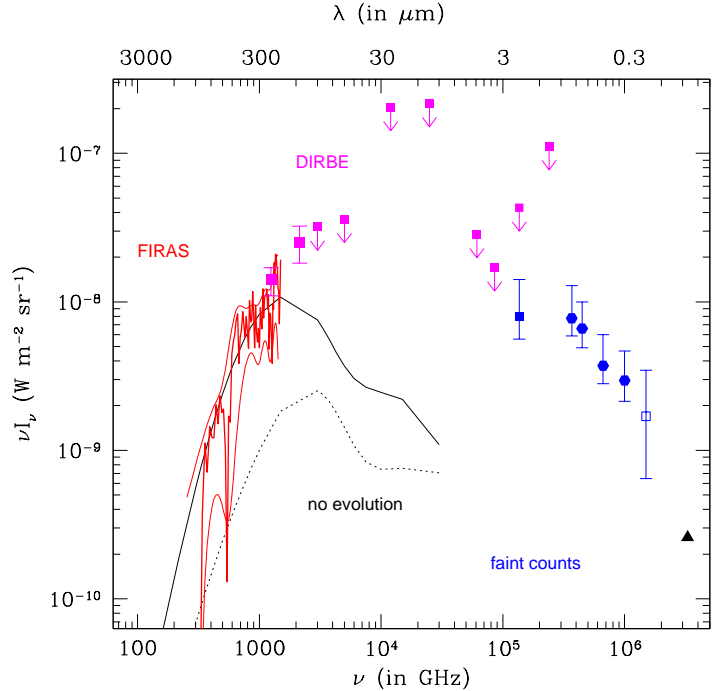,width=0.6\textwidth}
}
\caption{\small The cosmic optical and infrared backgrounds
(respectively COB and CIRB). The COB (solid square, solid dots,
and open square) is obtained from the faint 
counts and compiled by Pozzetti {\it et al.} (1998). The solid triangle gives
an upper limit by Vogel {\it et al.} (1995). The thick solid lines show
the CIRB extracted from FIRAS residuals in the cleanest regions 
of the sky (Puget {\it et al.} 1996; Guiderdoni {\it et al.} 1997). The thin
lines bracket the range that is consistent with the error bars. The 
solid squares give 
the upper limits and detections (at 140 and 240 $\mu$m) from DIRBE 
(Hauser {\it et al.} 1998). The no--evolution curve (dotted line)
is computed from the 
local IRAS luminosity function extrapolated to $z=8$, 
for $H_0=50$ km s$^{-1}$ 
Mpc$^{-1}$ and $\Omega_0=1$.  The solid line shows the best model 
(the so--called ``model E'') in Guiderdoni {\it et al.} (1998).}
\end{figure}

\section{The diffuse IR/submm background and submm counts}
A lower limit of the ``Cosmic Optical Background'' (hereafter COB) is 
currently estimated by summing up the faint counts obtained
in the Hubble Deep Field (HDF), and from ground--based observations. 
The shallowing of the slope suggests that the counts are close 
to convergence. 

In the submm range, Puget {\it et al.} (1996)
discovered an isotropic component in the FIRAS residuals
between 200 $\mu$m and 2 mm. This measure was confirmed 
by subsequent work in the cleanest regions of the sky (Guiderdoni {\it et al.}
1997), and by an independent determination (Fixsen {\it et al.} 1998).
The analysis of the DIRBE
dark sky has also led to the detection of the isotropic background at 240 and 
140 $\mu$m, and to upper limits at shorter wavelengths down to 2 $\mu$m
(Schlegel {\it et al.} 1998, Hauser {\it et al.} 1998). Recently, a measure
at 3.5 $\mu$m was proposed by Dwek \& Arendt 1998.
The results of these analyses seem in good agreement, though the exact level 
of the background around 200 $\mu$m is still a matter of debate. The 
controversy concerns the correction for the amount of Galactic dust in
the ionized gas uncorrelated with the HI gas. 

It appears very likely that this isotropic background is the long--sought
CIRB. As shown in fig. 2, its level 
is about 5--10 times the no--evolution prediction based on the local IR
luminosity function determined by IRAS. There is about twice as much flux
in the CIRB than in the COB. If the dust that emits at IR/submm wavelengths 
is mainly heated by young stellar populations, the sum of the fluxes of the 
CIRB and COB gives the level of the Cosmic Background associated to 
stellar nucleosynthesis (Partridge and Peebles 1967). The bolometric 
intensity (in W m$^{-2}$ sr$^{-1}$) is~:
\begin{equation}
I_{bol}=\int {\epsilon_{bol} \over 4\pi} {dl \over (1+z)^4} = {c\eta
\over 4\pi}{\rho_Z(z=0) \over (1+z_{eff})}
\end{equation}
where the physical emissivity due to young stars at cosmic time $t$ 
is $\epsilon(t)=\eta (1+z)^3 d\rho_Z(t)/dt$ and $z_{eff}$ is the effective 
redshift for stellar He and metal nucleosynthesis.
The census of the local density of heavy elements $\rho_Z(z=0)
\sim 1 \times 10^7$ $M_\odot$ Mpc$^{-3}$ gives an expected bolometric
intensity of the background $I_{bol} \simeq 50(1+z_{eff})^{-1}$ 
nW m$^{-2}$ sr$^{-1}$. This value is roughly consistent with the observations 
for $z_{eff} \sim 1$ -- 2.

Of course, it is not clear yet whether star formation is responsible
for the bulk of dust heating, or there is a significant contribution of AGNs. 
In order to address this issue, one has first to identify 
the sources that are responsible for the CIRB. At low $z$, 
it is well known that the IRAS satellite has discovered 
``luminous IR galaxies'' (hereafter LIRGs), mostly interacting systems, and 
the spectacular ``ultraluminous IR galaxies'' (hereafter ULIRGs), 
which are mergers and emit more than 95 \% of their
energy in the IR (see e.g. the review by Sanders and Mirabel 1996). 
The question of the origin of dust heating in these heavily--extinguished 
objects is a difficult one, because both starburst and AGN rejuvenation 
can be fueled by gas inflows triggered by interaction.
However, according to Genzel {\it et al.} (1998), 
the starburst generally contributes to 50--90 \% of the heating in local 
ULIRGs. Now, it is very likely that the 
high--redshift counterparts of the local LIRGs and ULIRGs are 
largely responsible 
for the CIRB. but the redshift evolution of the fraction and power of AGNs 
that are harbored in these distant objects is still unknown. 

Various submm surveys have been achieved or are in progress.
The FIRBACK program is a deep survey of 4 deg$^2$ at 175 $\mu$m with the 
ISOPHOT instrument aboard ISO. The analysis of about 1/4 of the Southern 
fields (that is, of 0.25 deg$^2$) unveils 24 sources (with $S_\nu > 100$ mJy), 
corresponding to a surface density five times larger 
than the no--evolution predictions based on the local IR luminosity 
function (Puget {\it et al.} 1998). The total catalogue of the 4 deg$^2$ will
include about 275 sources (Dole {\it et al.} 1999). The radio
and optical follow--up for identification is still in progress. 
This strong evolution is confirmed
by the other 175 $\mu$m deep survey by Kawara {\it et al.} (1998). 
Various deep surveys
at 850 $\mu$m have been achieved with the SCUBA instrument at the JCMT
(Smail {\it et al.} 1997, Hughes {\it et al.} 1998, Barger {\it et al.} 1998, 
Eales {\it et al.} 1998). They also unveil a surface density of sources
(with $S_\nu > 2$ mJy) much larger than the no--evolution predictions
(by two or three orders of magnitude !).
The total number of sources so far discovered in SCUBA deep surveys 
now reaches about 40 (see e.g. Blain {\it et al.} 1998).
The tentative optical identifications seem to show that some of
these objects look like distant ULIRGs (Smail {\it et al.} 1998, Lilly 
{\it et al.} 1999). 
In the HDF, 4 of the brightest 5 sources seem to lie
between redshifts 2 and 4 (Hughes {\it et al.} 1998), but the 
optical identifications are still a matter of debate 
(Richards, 1998). The source SMM 02399-0136 at $z=2.803$, which is
gravitationally amplified by the foreground cluster A370, 
is clearly an AGN/starburst galaxy 
(Ivison {\it et al.} 1998, Frayer {\it et al.} 1998).
Fig. 3 gives an account of the faint counts in the submm range.

\begin{figure}[htb]
\centerline{
\psfig{figure=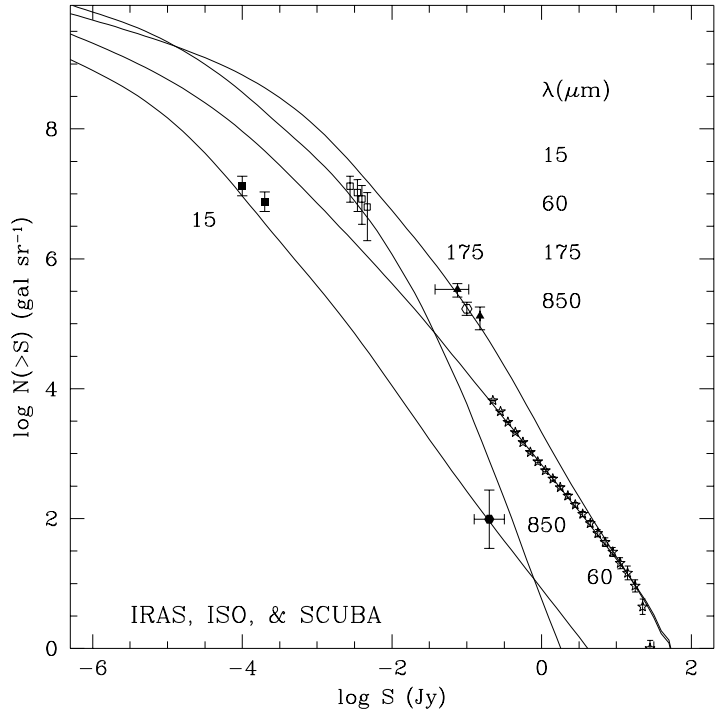,width=0.6\textwidth}
}
\caption{\small Predictions for faint counts at 15 $\mu$m, 
60 $\mu$m, 175 $\mu$m, and 850 $\mu$m. These predictions
correspond to model E and have been obtained before the data, on the basis of 
the CIRB and IRAS counts. Open stars: 
Faint Source Survey (Lonsdale {\it et al.} 1990). Solid hexagon: 
Rush {\it et al.} (1993). Solid squares:
ISO--HDF with ISOCAM at 15 $\mu$m (Oliver {\it et al.} 1997). Open
hexagon: ISOPHOT at 175 $\mu$m (Kawara {\it et al.} 1998). Solid
triangles: one of the Southern fields of the ISOPHOT FIRBACK survey (Puget
{\it et al.} 1998). Open squares : deep SCUBA survey (Smail {\it et al.}
1997).}
\end{figure}

\section{Modeling dust spectra in a semi--analytic framework}
Various models have been proposed to account for the FIR/submm emission of 
galaxies and to predict forthcoming observations. The level of 
sophistication (and complexity) increases from pure luminosity and/or density 
evolution extrapolated from the IRAS local luminosity function with $(1+z)^n$ 
laws, and modified black--body spectra, to physically--motivated spectral 
evolution.

Guiderdoni {\it et al.} (1998) proposed a consistent modeling of IR/submm
galaxy counts in the paradigm of hierarchical clustering. Only stellar
heating is taken into account.
The IR/submm spectra of galaxies are computed in the following way:
(i) follow chemical evolution of the gas;
(ii) implement extinction curves which depend 
on metallicity as in the Milky Way, the LMC and SMC;
(iii) compute $\tau_\lambda$;
(iv)  assume the so--called ``slab'' geometry where the star and dust 
components are homogeneously mixed with equal height scales. 
(v) compute a spectral energy distribution by assuming a mix of 
various dust components. The contributions are fixed in order to reproduce the 
observational correlation of IRAS colours with total IR luminosity
(Soifer \& Neugebauer 1991).

\begin{figure}[htb]
\centerline{
\psfig{figure=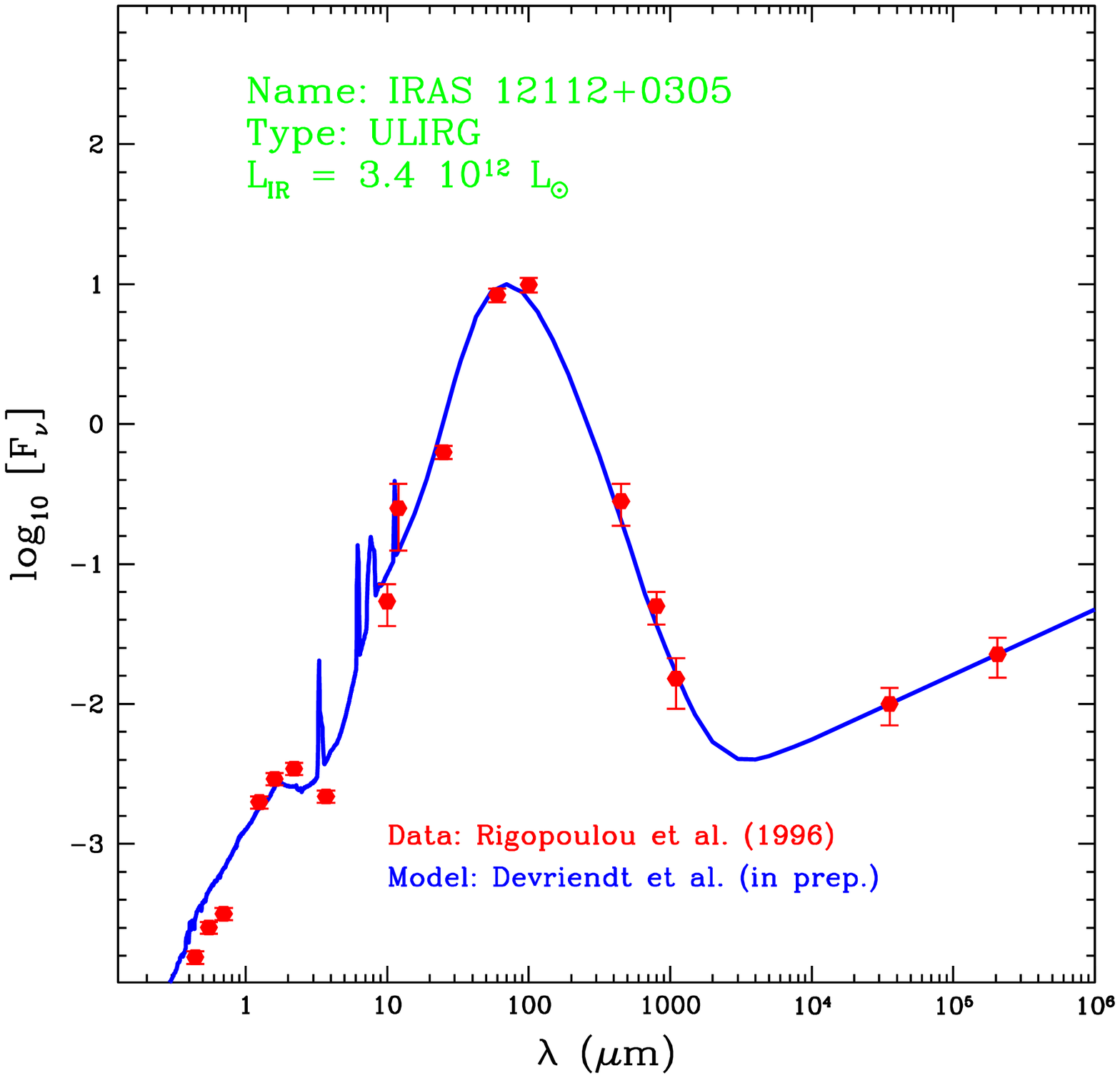,width=0.6\textwidth}
}
\caption{\small Spectral energy distribution of a typical ULIRG. The solid
line shows the fit of the data with a theoretical spectrum computed by
Devriendt {\it et al.} (1999).}
\end{figure}

These FIR/submm spectra are implemented in a semi--analytic model of galaxy
formation and evolution. This type of model has been very effective in 
computing the optical properties of galaxies in the paradigm of hierarchical
clustering. We only extend this approach to the IR/submm range, and take
the standard CDM case with $H_0$=50 km s$^{-1}$ Mpc$^{-1}$, 
$\Omega_0=1$, $\Lambda=0$, and $\sigma_8=0.67$.
We assume a Star Formation Rate $SFR(t)=M_{gas}/t_*$, with $t_* \equiv \beta
t_{dyn}$ and a Salpeter IMF ($x=1.35$). The efficiency parameter 
$1/\beta =0.01$ gives a nice fit of
local spirals. The robust result of this type of modeling is a
cosmic star formation rate history that is too flat with respect to the data. 

As a phenomenological way of
reproducing the steep rise of the cosmic SFR history from $z=0$ to $z=1$, 
we introduce a ``burst'' mode of star formation 
involving a mass fraction that increases with $z$
as $(1+z)^4$, with ten times higher efficiencies $1/\beta=0.1$.
In order to reproduce the level of the CIRB, we have to assume that a small 
fraction of the gas mass (typically less than 10 \%) is involved in
star formation with a top--heavy IMF in heavily--extinguished objects 
(ULIRG--type galaxies).
The so--called ``model E'' with these assumptions fairly reproduces the cosmic 
SFR and luminosity densities, as well as the CIRB (see fig. 1 and 2).

Fig. 3 gives the predictions of number counts at 15, 60, 175, and 850 $\mu$m
for this model.
The agreement of the predictions with the data seems good enough 
to suggest that these counts do probe the evolving 
population contributing to the CIRB. The model
shows that 15 \%  and 60 \% of the CIRB respectively at 175 $\mu$m 
and 850 $\mu$m are built up by objects brighter than
the current limits of ISOPHOT and
SCUBA deep fields. The predicted median redshift of the ISO--HDF
is $z \sim 0.8$. It increases to $z \sim 1.2$ for the deep ISOPHOT
surveys, and to $z \ge 2$ for SCUBA, though the latter value is
very sensitive to the details of the evolution. 

An extension of the spectra and counts to the near--IR, optical and 
ultraviolet ranges is in progress (Devriendt {\it et al.} 1999, 
Devriendt \& Guiderdoni 1999). A fit of a typical ULIRG is proposed in fig.
4 as an example of what can be obtained with these extended spectra.

\begin{figure}[htb]
\centerline{ 
\psfig{figure=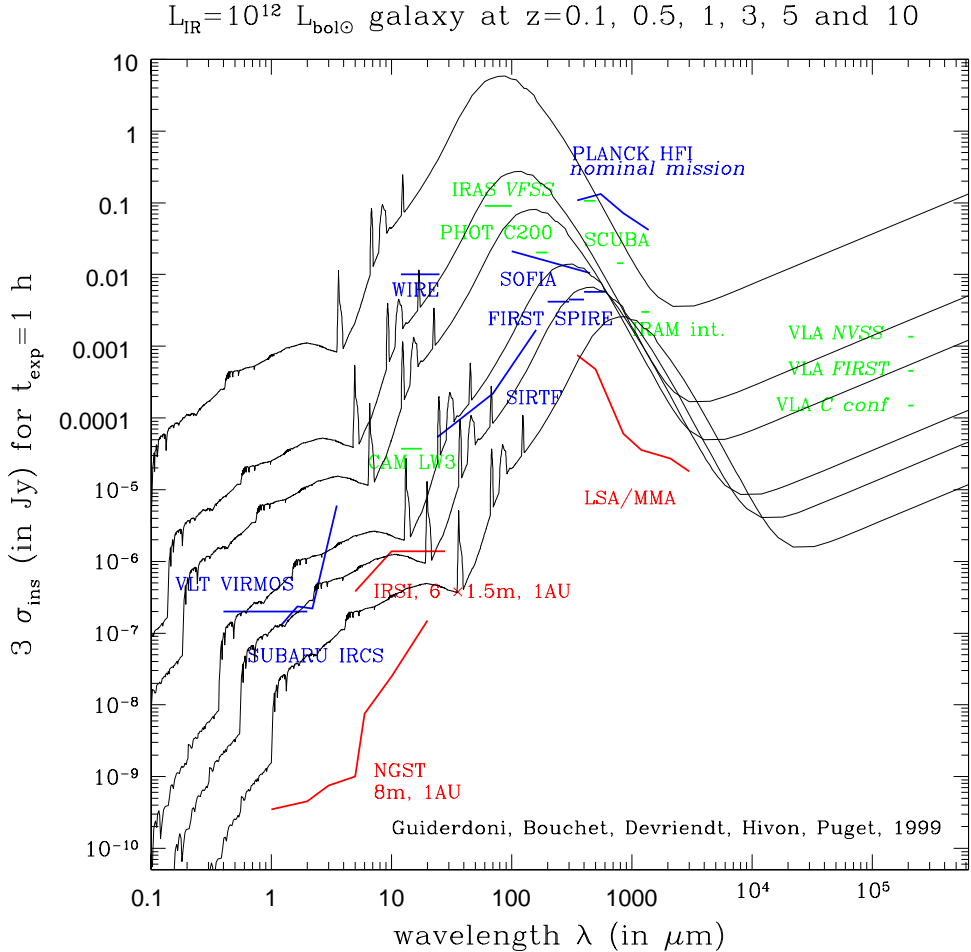,width=0.8\textwidth}
}
\caption{\small Observer--frame model spectra of a $L_{IR}=10^{12} L_{\odot}$
ULIRG at increasing redshifts (from top to bottom).
The reader in invited to note that the apparent flux in 
the submm range is almost insensitive to redshift, because the shift of the
100 $\mu$m bump counterbalances the distance dimming. The instrumental 
sensitivities of various past, current, and forthcoming instruments 
(ground--based and satellite--borne telescopes)
are plotted. This figure is available upon request to {\tt guider@iap.fr}.}
\end{figure}

\section{Future instruments}
Fig. 5 gives the far--UV to submm spectral energy distribution that is
typical of a 
$L_{IR}=10^{12}$  $L_{bol\odot}$ ULIRG at various redshifts. 
This model spectrum is taken from the computation of 
Devriendt {\it et al.} (1999).
The reader should note the specific behavior of the observed flux at
submm wavelengths, where the shift of the 60 -- 100 $\mu$m 
rest--frame emission bump
counterbalances distance dimming. The instrumental sensitivities
of various past and on--going satellite and ground--based
instruments are plotted on this
diagram~: the IRAS  {\it Very Faint Source Survey} at 60 $\mu$m,
ISOCAM at 15 $\mu$m, ISOPHOT at 175 $\mu$m, the IRAM interferometer
at 1.3 mm, SCUBA at 450 and 850 $\mu$m, and various surveys with the VLA.
Forthcoming missions and facilities include WIRE, SIRTF, SOFIA, 
the PLANCK {\it High Frequency 
Instrument}, the FIRST {\it Spectral and Photometric Imaging REceiver},
and the imaging modes of the SUBARU IRCS and VLT VIRMOS instruments.
Finally, the capabilities of the NGST, MMA/LSA and Infrared Space 
Interferometer (DARWIN) are also plotted.

The final sensitivity of the next--generation instruments
observing at FIR and submm wavelengths (WIRE, SIRTF, SOFIA,
PLANCK, FIRST) is going to be confusion limited. However,
the observation of a large sample of ULIRG--like objects in the redshift
range 1--5 should be possible. More specifically,
the all--sky shallow
survey of PLANCK {\it HFI}, and the medium--deep survey of FIRST {\it SPIRE}
(to be launched by ESA in 2007), will respectively produce bright
($S_\nu >$ a few 100 mJy) and faint ($S_\nu >$ a few 10 mJy) counts that 
will be complementary. A 10 deg$^2$ survey with {\it SPIRE} will
result in $\sim 10^4$ sources.
The study of the $250/350$ and $350/500$ colors are suited to point out 
sources which are likely to be at high redshifts. These sources can be 
eventually followed at 100 and 170 $\mu$m by the FIRST 
{\it Photoconductor Array Camera \& Spectrometer} and by the FTS mode
of {\it SPIRE}, to get the spectral energy distribution at $200 \leq \lambda 
\leq 600$ $\mu$m with a typical resolution $R\equiv \lambda /\Delta\lambda=20$.
After a photometric and spectroscopic followup, the submm observations should
readily probe the bulk of (rest--frame IR) luminosity associated with
star formation. The reconstruction of the cosmic SFR comoving density will 
thus take into account the correct luminosity budget of high--redshift 
galaxies.
However, the spatial resolution of the submm instruments will be limited, 
and only the MMA/LSA 
should be able to resolve the FIR/submm sources and study the details of their 
structure.

\section{Conclusions}
\begin{enumerate}
\item There is now strong evidence that high--redshift galaxies emit much 
more IR luminosity than predictions based on the local IR luminosity 
function, without evolution. The submm counts seem to unveil the 
bright end of the population that is responsible for the CIRB.
\item The issue of the relative contributions of the starbursts and AGNs to
dust heating is still unsolved. Local ULIRGs seem to be dominated by 
starburst heating, but the behavior at higher redshift is unknown. 
\item It is difficult to correct for
the influence of dust on the basis of the optical spectra alone.
Multi--wavelength studies are clearly necessary to address the 
history of the cosmic SFR density, through a correct assessment of the 
luminosity budget.
\item Under the assumption that starburst heating is dominant, simple models 
in the paradigm of hierarchical clustering do reproduce the current 
IR/submm data. 
\item The current studies on faint counts at submm wavelengths will guide
models for the preparation of the observing strategies with forthcoming 
instruments : e.g., SIRTF, SOFIA, the PLANCK {\it High Frequency 
Instrument}, the FIRST {\it Spectral and Photometric Imaging REceiver},
and the MMA/LSA. A large number of high--redshift sources should be 
observable with these IR/submm instruments.
\end{enumerate}

\vfill
\end{document}